\documentclass[11pt]{revtex4}    % Specifies the document style.
\usepackage{amssymb,epsf}
\usepackage{latexsym}
\begin{document}

\title{Topological Black Holes in Gauss-Bonnet Gravity
with conformally invariant Maxwell source}
\author{S. H. Hendi\footnote{hendi@mail.yu.ac.ir}}

\address{Physics Department,
College of Sciences, Yasouj University, Yasouj
75914, Iran\\
Research Institute for Astrophysics and Astronomy of Maragha
(RIAAM), P.O. Box 55134-441, Maragha, Iran}

\begin{abstract}
In this paper, we present a class of rotating solutions in
Gauss--Bonnet gravity in the presence of cosmological constant and
conformally invariant Maxwell field and study the effects of the
nonlinearity of the Maxwell source on the properties of the
spacetimes. These solutions may be interpret as black brane
solutions with inner and outer event horizons provide that the
mass parameter $m$ is greater than an extremal value $m_{ext}$, an
extreme black brane if $m=m_{ext}$ and a naked singularity
otherwise. We investigate the conserved and thermodynamics
quantities for asymptotically flat and asymptotically $AdS$ with
flat horizon. We also show that the conserved and thermodynamic
quantities of these solutions satisfy the first law of
thermodynamics.
\end{abstract}

\pacs{04.40.Nr, 04.20.Jb, 04.70.Bw, 04.70.Dy} \maketitle

\section{Introduction}
The presence of higher curvature terms can be seen in many
theories such as renormalization of quantum field theory in curved spacetime \cite{Birrel}%
, or in construction of low energy effective action of string theory \cite%
{Green} and so on. These examples motivate one to consider the
more general class of gravitational action
\[
I_{G}=\frac{1}{16\pi }\int d^{n+1}x\sqrt{-g}\mathcal{F}(R,R_{\mu
\nu }R^{\mu \nu },R_{\mu \nu \rho \sigma }R^{\mu \nu \rho \sigma
}),
\]
where $R$, $R_{\mu \nu }$ and $R_{\mu \nu \rho \sigma }$ are Ricci
scalar, Ricci and Riemann tensor respectively. Among the higher
curvature gravity theories, the so called Lovelock gravity is
quite special, whose Lagrangian consist of the dimensionally
extended Euler densities. This Lagrangian is obtained by Lovelock
as he tried to calculate the most general tensor that
satisfies properties of Einstein's tensor in higher dimensions \cite%
{Lovelock}. Since the Lovelock tensor does not contain derivatives
of metrics of order higher than second, the quantization of
linearized Lovelock theory is free of ghosts \cite{Boulware}. The
gravitational action of Lovelock theory can be written as
\cite{Lovelock}
\begin{equation}
I_{G}=\int d^{d}x\sqrt{-g}\sum_{k=0}^{[d/2]}\alpha
_{k}\mathcal{L}_{k},
\end{equation}%
where $[z]$ denotes integer part of $z$, $\alpha _{k}$ is an
arbitrary constant and $\mathcal{L}_{k}$ is the Euler density of a
$2k$-dimensional manifold,
\begin{equation}
\mathcal{L}_{k}=\frac{1}{2^{k}}\delta _{\rho _{1}\sigma _{1}\cdots
\rho _{k}\sigma _{k}}^{\mu _{1}\nu _{1}\cdots \mu _{k}\nu
_{k}}R_{\mu _{1}\nu
_{1}}^{\phantom{\mu_1\nu_1}{\rho_1\sigma_1}}\cdots R_{\mu _{k}\nu _{k}}^{%
\phantom{\mu_k \nu_k}{\rho_k \sigma_k}}.  \label{Lov2}
\end{equation}%
In Eq. (\ref{Lov2}) $\delta _{\rho _{1}\sigma _{1}\cdots \rho
_{k}\sigma _{k}}^{\mu _{1}\nu _{1}\cdots \mu _{k}\nu _{k}}$ is the
generalized totally anti-symmetric Kronecker delta. It is
worthwhile to mention that in $d$ dimensions, all terms for which
$k>[d/2]$ are identically equal to zero, and the term $k=d/2$ is a
topological term. So, only terms for which $k<d/2$ are
contributing to the field equations. In this paper we want to
restrict ourself to the first three terms of Lovelock gravity. The
first term is the cosmological term and the second and third terms
are the Einstein and second order Lovelock (Gauss-Bonnet) terms
respectively.

In the last decades a renewed interest appears in Lovelock
gravity. In particular, exact static spherically symmetric black
hole solutions of the Gauss-Bonnet gravity have been found in Ref.
\cite{Boulware,Astefanesei}. Exact solutions of static and
rotating of the Maxwell-Gauss-Bonnet and Born-Infeld-Gauss-Bonnet
models have been investigated in Ref.
\cite{Wil1,DH,Deh1,Deh2,DehMagGB,Deh3}. The thermodynamics of the
uncharged static spherically black hole solutions has been
considered in \cite{MS}. Properties of solutions with nontrivial
topology have been studied in \cite{Cai,NUT} and charged black
hole solutions have been found in \cite{Wil1,Od1}.

In the conventional, straightforward generalization of the Maxwell
field to higher dimensions one essential property of the
electromagnetic field is lost, namely, conformal invariance. The
first black hole solution derived
for which the matter source is conformally invariant is the Reissner-Nordstr%
\"{o}m solution in four dimensions. Indeed, in this case the
source is given by the Maxwell action which enjoys the conformal
invariance in four dimensions. Massless spin-$1/2$ fields have
vanishing classical stress tensor trace in any dimension, while
scalars can be \textquotedblleft improved\textquotedblright\ to
achieve $T_{\alpha }^{\alpha }=0$, thereby guaranteeing invariance
under the special conformal (or full Weyl) group, in accord with
their scale-independence \cite{DesSch,EspSto,CodOsb}. Maxwell
theory can be studied in a gauge which is invariant under
conformal
rescalings of the metric, and firstly, has been proposed by Eastwood and Singer \cite%
{EasSin}. Also, Poplawski \cite{Pop} have been showed the
equivalence between the Ferraris--Kijowski and Maxwell Lagrangian
results from the invariance of the latter under conformal
transformations of the metric tensor. Quantized Maxwell theory in
a conformally invariant gauge have been investigated by Esposito
\cite{Esp}. Also, there exists a conformally invariant extension
of the Maxwell action in higher dimensions (Generalized Maxwell
Field, GMF), if one uses the lagrangian of the $U(1)$ gauge field
in the form \cite{HasMar}
\begin{equation}
I_{GMF}=\kappa \int d^{d}x\sqrt{-g}\left( F_{\mu \nu }F^{\mu \nu
}\right) ^{d/4},  \label{IGMF}
\end{equation}
where $F_{\mu \nu }=\partial _{\mu }A_{\nu }-\partial _{\nu
}A_{\mu }$ is the \ Maxwell tensor and $\kappa $ is an arbitrary
constant. It is straightforward to show that the action
(\ref{IGMF}) is invariant under conformal transformation ($g_{\mu
\nu }\longrightarrow \Omega ^{2}g_{\mu \nu
}$ and $A_{\mu }\longrightarrow A_{\mu }$) and for $d=4$, the action (\ref%
{IGMF}) reduces to the Maxwell action as it should be. The
energy-momentum tensor associated to $I_{GMF}$ is given by
\begin{equation}
T_{\mu \nu }=\kappa \left( dF_{\mu \rho }F_{\nu }^{~\rho
}F^{(d/4)-1}-g_{\mu \nu }F^{d/4}\right) ,  \label{T}
\end{equation}
where $F=F_{\mu \nu }F^{\mu \nu }$ and it is easy to show that
$T_{\mu }^{\mu }=0$. In what follows, we consider the action
(\ref{IGMF}) as the matter source coupled to the Gauss-Bonnet
gravity. The idea is to take advantage of the conformal symmetry
to construct the analogues of the four-dimensional
Reissner-Nordstr\"{o}m black hole solutions in higher dimensions.
The form of the energy-momentum tensor (\ref{T}), automatically
restricts the dimensions to be only multiples of four.

The main aim of this work is to present analytical solutions for
typical class of rotating spacetime in Gauss-Bonnet theory coupled
to conformally invariant Maxwell source with negative cosmological
constant. As we show later, these solutions have some interesting
properties, specially in the electromagnetic fields, which do not
occur in Gauss-Bonnet gravity in the present of ordinary Maxwell
field. In this paper we discuss black hole solutions of
Gauss-Bonnet-conformally invariant Maxwell source and investigate
their properties and calculate the conserved and thermodynamics
quantities.

The outline of our paper is as follows. In next Section, we
present the basic field equations and general formalism of
calculating the conserved quantities. In section \ref{black hole},
we present the topological black hole of Gauss-Bonnet gravity in
the presence of conformally invariant Maxwell source. Then, we
calculate the thermodynamic quantities of asymptotically flat
solutions and investigate the first law of thermodynamics. In next
subsection \ref{rotating} we introduce the rotating solutions with
flat horizon and compute the thermodynamic and conserved
quantities of them. We also perform a stability analysis of this
solutions both in canonical and grand canonical ensemble. Finally,
we finish our paper with some concluding remarks.

\section{Field Equations in Gauss-Bonnet Gravity with Conformally Invariant
Maxwell Source\label{Fiel}} The action of Gauss-Bonnet gravity in
the presence of conformally invariant electromagnetic field is
\begin{eqnarray}
I_{G} &=&-\frac{1}{16\pi
}\int_{\mathcal{M}}d^{4p+4}x\sqrt{-g}\left\{ R-2\Lambda +\alpha
(R_{\mu \nu \gamma \delta }R^{\mu \nu \gamma \delta }-4R_{\mu \nu
}R^{\mu \nu }+R^{2})-\kappa (F_{\mu \nu }F^{\mu \nu
})^{p+1}\right\}   \nonumber \\
&&-\frac{1}{8\pi }\int_{\partial
\mathcal{M}}d^{4p+3}x\sqrt{-\gamma }\left\{ K+2\alpha \left(
J-2\widehat{G}_{ab}K^{ab}\right) \right\} ,  \label{Ig}
\end{eqnarray}%
where $\Lambda =-(2p+1)(4p+3)/l^{2}$ is the negative cosmological
constant for asymptotically AdS solutions and $\alpha $ is the
Gauss-Bonnet coefficient with dimension
$(\mathrm{length})^{2}$\textbf{.} The second integral in Eq.
(\ref{Ig}) is the Gobbons-Hawking surface term and its counterpart
for the Gauss-Bonnet gravity which is chosen such that the
variational principle is well defined \cite{MyeDavis}. In this
term, $\gamma _{ab}$ is induced metric
on the boundary $\partial \mathcal{M}$, $K$ is trace of extrinsic curvature $%
K^{ab}$ of the boundary, $\widehat{G}_{ab}(\gamma )$ is Einstein
tensor of the metric $\gamma _{ab}$, and $J$ is trace of the
tensor
\begin{equation}
J_{ab}=\frac{1}{3}%
(K_{cd}K^{cd}K_{ab}+2KK_{ac}K_{b}^{c}-2K_{ac}K^{cd}K_{db}-K^{2}K_{ab}).
\label{Jab}
\end{equation}%
Varying the action with respect to the metric tensor $g_{\mu \nu
}$ and electromagnetic field $A_{\mu }$ the equations of
gravitation and electromagnetic fields are obtained as
\begin{equation}
G_{\mu \nu }+\Lambda g_{\mu \nu }-\alpha H_{\mu \nu }=2\kappa
\left[
(p+1)F_{\mu \rho }F_{\nu }^{~\rho }F^{p}-\frac{1}{4}g_{\mu \nu }F^{p+1}%
\right] ,  \label{Geq}
\end{equation}%
\begin{equation}
\partial _{\mu }\left( \sqrt{-g}F^{\mu \nu }F^{p}\right) =0,  \label{Maxeq}
\end{equation}%
where $G_{\mu \nu }$ is the Einstein tensor and $H_{\mu \nu }$\ is
the divergence-free symmetric tensor
\begin{eqnarray}
&&H_{\mu \nu }=4R^{\rho \sigma }R_{\mu \rho \nu \sigma }-2R_{\mu
}^{\ \rho \sigma \lambda }R_{\nu \rho \sigma \lambda }-2RR_{\mu
\nu }+4R_{\mu \lambda
}R_{\text{ \ }\nu }^{\lambda }  \nonumber \\
&&+\frac{1}{2}g_{\mu \nu }(R_{\kappa \lambda \rho \sigma
}R^{\kappa \lambda \rho \sigma }-4R_{\rho \sigma }R^{\rho \sigma
}+R^{2}).  \label{Heq}
\end{eqnarray}
Hereafter we set $\kappa =(-1)^{p}$ without loss of generality and
consequently the energy density (the $T_{\widehat{0}\widehat{0}}$
component of the energy-momentum tensor in the orthonormal frame)
is positive.

Equation (\ref{Geq}) does not contain the derivative of the
curvatures, and therefore the derivatives of the metric higher
than two do not appear. In general the action $I_{G}$ is diverged
when evaluated on solutions, as is
the Hamiltonian and other associated conserved charges. For asymptotically $%
AdS$ solutions, one can instead deal with these divergences via
the counterterm method inspired by $AdS/CFT$ correspondence
\cite{Mal}. In the present context this correspondence furnishes a
means for calculating the action and conserved quantities
intrinsically by adding additional terms on the boundary that are
curvature invariants of the induced metric. Although there may
exist a very large number of possible invariants one could add in
a given dimension, only a finite number of them are non vanishing
as the boundary is taken to infinity. Its many applications
include computations of conserved quantities for black holes with
rotation, various topologies, rotating black strings with zero
curvature horizons and rotating higher genus black branes
\cite{Deh33}. Usually, the counterterm method applies for the case
of a specially infinite boundary, but it was also employed for the
computation of the conserved and thermodynamic quantities in the
case of a finite boundary \cite{DM1}. Thus, in order to extract
the physically relevant information, the on-shell action has to be
renormalized by adding counterterms, which cancel the infinities
of $I_{G}$ in the absence of matter. This counterterms is
\cite{Kraus}
\begin{eqnarray}
I_{\mathrm{ct}} &=&\frac{1}{8\pi }\int_{\partial
\mathcal{M}_{\infty
}}d^{4p+3}x\sqrt{-\gamma }\{\frac{4p+2}{l}-\frac{l\Theta (2p)}{2(4p+1)}R-%
\frac{l^{3}\Theta (2p-1)}{2(4p-1)(4p+1)^{2}}\left( R_{ab}R^{ab}-\frac{%
(4p+3)R^{2}}{8(2p+1)}\right)   \nonumber \\
&&\ +\frac{l^{5}\Theta (2p-2)}{(4p+1)^{3}(4p-1)(4p-3)}[\frac{12p+11}{4(4p+2)}%
RR_{ab}R^{ab}-\frac{(4p+3)(4p+5)}{16(4p+2)^{2}}R^{3}  \nonumber \\
&&\ -2R^{ab}R_{acbd}R^{cd}+\frac{4p+1}{2(4p+2)}R^{ab}\nabla
_{a}\nabla _{b}R-R^{ab}\square R_{ab}+\frac{R\square
R}{2(4p+2)}]\}+...,  \label{Ict}
\end{eqnarray}%
where $\Theta (x)$ is the step function which is equal to one for
$x\geq 0$ and zero otherwise. Thus, the total finite action can be
written as a linear combination of the gravity term $I_{G}$ and
the counterterm $I_{ct}$. Having the total finite action, one can
use the Brown and York definition of energy-momentum tensor
\cite{BY} to construct a divergence free stress-energy tensor.
This tensor is
\begin{eqnarray}
T^{ab} &=&\frac{1}{8\pi }\{(K^{ab}-K\gamma ^{ab})-\frac{4p+2}{l}\gamma ^{ab}+%
\frac{l}{4p+1}(R^{ab}-\frac{1}{2}R\gamma ^{ab})  \nonumber \\
&&\ \ +\frac{l^{3}\Theta
(2p-1)}{(4p-1)(4p+1)^{2}}[-\frac{1}{2}\gamma
^{ab}(R^{cd}R_{cd}-\frac{4p+3}{4(4p+2)}R^{2})-\frac{4p+3}{4(2p+1)}RR^{ab}
\nonumber \\
&&\ \ +2R_{cd}R^{acbd}-\frac{4p+1}{2(4p+2)}\nabla ^{a}\nabla
^{b}R+\nabla ^{2}R^{ab}-\frac{1}{2(4p+2)}\gamma ^{ab}\nabla
^{2}R]+...\}.  \label{Stres}
\end{eqnarray}%
When there is a Killing vector field $\mathcal{\xi }^{a}$ on the
boundary, the quasilocal conserved quantities associated with the
stress tensors of Eq. (\ref{Stres}) can be written as
\begin{equation}
\mathcal{Q}(\mathcal{\xi )}=\int_{\mathcal{B}}d^{4p+2}\varphi \sqrt{\sigma }%
T_{ab}n^{a}\mathcal{\xi }^{b},  \label{quasi}
\end{equation}%
where $\sigma $ is the determinant of the metric $\sigma _{ij}$,
and $n^{a}$ is the timelike unit normal vector to the boundary
$\mathcal{B}$\textbf{.}
In the context of counterterm method, the limit in which the boundary $%
\mathcal{B}$ becomes infinite ($\mathcal{B}_{\infty }$) is taken,
and the counterterm prescription ensures that the action and
conserved charges are finite.

\section{Black Hole Solutions:\label{black hole}}

\subsection{Topological Black Holes}
Here we want to obtain the $(4p+4)$-dimensional static solutions of Eqs. (%
\ref{Geq}) and (\ref{Maxeq}). We assume that the metric has the
following form:
\begin{equation}
ds^{2}=-f(\rho )dt^{2}+\frac{d\rho ^{2}}{f(\rho )}+\rho
^{2}d\Omega _{k}^{2}, \label{met}
\end{equation}
where
\begin{equation}
d\Omega _{k}^{2}=\left\{
\begin{array}{cc}
d\theta
_{1}^{2}+\sum\limits_{i=2}^{4p+2}\prod\limits_{j=1}^{i-1}\sin
^{2}\theta _{j}d\theta _{i}^{2} & k=1 \\
d\theta _{1}^{2}+\sinh ^{2}\theta _{1}d\theta _{2}^{2}+\sinh
^{2}\theta
_{1}\sum\limits_{i=3}^{4p+2}\prod\limits_{j=2}^{i-1}\sin
^{2}\theta
_{j}d\theta _{i}^{2} & k=-1 \\
\sum\limits_{i=1}^{4p+2}d\phi _{i}^{2} & k=0%
\end{array}
\right. ,  \label{dOmega}
\end{equation}
which represents the line element of an $(4p+2)$-dimensional
hypersurface with constant curvature $(4p+1)(4p+2)k$ and volume
$V_{4p+2}$.

Using Eq. (\ref{Maxeq}), one can show that the vector potential, in $(4p+4)$%
-dimensions, can be written as
\begin{equation}
A_{\mu }=-\frac{q}{\rho }\delta _{\mu }^{0},  \label{Amu1}
\end{equation}%
where $q$ is an integration constant which is related to the
charge parameter. The conformally invariant Maxwell equation
implies that the electric field in $(4p+4)$-dimensions is
proportional to $\rho ^{-2}$ and
given by%
\begin{equation}
F_{tr}=\frac{q}{\rho ^{2}}.
\end{equation}
One may show that the metric function
\begin{equation}
f(\rho )=k+\frac{\rho ^{2}}{8p(4p+1)\alpha }\left( 1-\sqrt{1-g(\rho )}%
\right) {,}  \label{fr}
\end{equation}
where
\[
g(\rho )=16p(4p+1)\alpha \left( -\frac{\Lambda
}{(2p+1)(4p+3)}-\frac{m}{\rho ^{4p+3}}+\frac{2^{p}q^{2p+2}}{\rho
^{4p+4}}\right) {,}
\]%
satisfies the field equations (\ref{Geq}), where $m$ is the mass
parameter. The metric function $f(\rho )$ is real in the whole
range $0\leq \rho <\infty $ for uncharged solution ($q=0$),
provided that $\alpha \leqslant l^{2}/[16p(4p+1)]$. For charged
real solution one needs a transformation to make them real
\cite{DH,DehBord}. In the other word, $f(\rho )$ is real\ only in
the range $r_{0}\leqslant \rho <\infty $, where $r_{0}$ is the
largest real root of the following equation%
\begin{equation}
\left( \frac{1}{16p(4p+1)\alpha }+\frac{\Lambda
}{(2p+1)(4p+3)}\right) r_{0}^{4p+4}+mr_{0}-2^{p}q^{2p+2}=0{.}
\end{equation}%
\bigskip In order to restrict spacetime to the region $\rho \geqslant r_{0}$%
, we introduce a new radial coordinate $r$ such as \cite{DH}
\begin{equation}
r^{2}=\rho ^{2}-r_{0}^{2}\Longrightarrow d\rho ^{2}=\frac{r^{2}}{%
r^{2}+r_{0}^{2}}dr^{2}{.}  \label{ChangeVar}
\end{equation}%
With this new coordinate, the above metric becomes%
\begin{equation}
ds^{2}=-f(r)dt^{2}+\frac{r^{2}dr^{2}}{\left( r^{2}+r_{0}^{2}\right) f(r)}%
+\left( r^{2}+r_{0}^{2}\right) ^{2}d\Omega _{k}^{2},
\label{metric}
\end{equation}%
where $d\Omega _{k}^{2}$ is the same as Eq. (\ref{dOmega}) and the
functions
$A_{\mu }$\ and $f(r)$ change to%
\begin{equation}
A_{\mu }=\frac{-q}{\sqrt{r^{2}+r_{0}^{2}}}\delta _{\mu }^{0},
\end{equation}%
\begin{equation}
f(r)=k+\frac{r^{2}+r_{0}^{2}}{8p(4p+1)\alpha }\left( 1-\sqrt{1-g(r)}\right) {%
,}  \label{F(r)}
\end{equation}%
\begin{equation}
g(r)=16p(4p+1)\alpha \left( -\frac{\Lambda
}{(2p+1)(4p+3)}-\frac{m}{\left( r^{2}+r_{0}^{2}\right) ^{\left(
4p+3\right) /2}}+\frac{2^{p}q^{2p+2}}{\left(
r^{2}+r_{0}^{2}\right) ^{2p+2}}\right) {.}  \label{g(r)}
\end{equation}%
In order to consider the asymptotic behavior of the solution, we
put $m=q=0$ where the metric function reduces to
\begin{equation}
f(r)=k+\frac{r^{2}+r_{0}^{2}}{8p(4p+1)\alpha }\left( 1-\sqrt{1+\frac{%
16p(4p+1)\alpha \Lambda }{(2p+1)(4p+3)}}\right) .  \label{Fg0}
\end{equation}%
Equation (\ref{Fg0}) shows that the asymptotic behavior of the
solution is AdS or dS provided $\Lambda <0$ or $\Lambda >0$. The
case of asymptotic flat solutions ($\Lambda =0$) is permitted only
for $k=1$. It easy to show that
in the vicinity of $r=0$%
\[
R_{\alpha \beta \gamma \delta }R^{\alpha \beta \gamma \delta
}\propto r^{-4},
\]%
and thus, the metric given by Eqs. (\ref{metric}), (\ref{F(r)})
and (\ref{g(r)}) has an essential timelike singularity at $r=0$.
Seeking possible black hole solutions, we turn to looking for the
existence of horizons. The event
horizon(s), if there exists any, is (are) located at the root(s) of $%
g^{rr}=f(r)=0$. Denoting the largest real root of $f(r)$ by
$r_{+}$, we
consider first the case that $f(r)$ has only one real root (see Fig. \ref%
{Fig f(r)}, continuous line). In this case $f(r)$ is minimum at
$r_{+}$ and therefore $f^{^{\prime }}(r_{+})=0$. That is,
\begin{equation}
(4p+1)k\left[ (r_{+}^{2}+r_{0}^{2})^{2}+4\alpha pk(4p-1)\right]
(r_{+}^{2}+r_{0}^{2})^{4p}-\frac{\Lambda }{(2p+1)}%
(r_{+}^{2}+r_{0}^{2})^{4p+4}-2^{p}q^{2p+2}=0.  \label{f'}
\end{equation}

\begin{figure}[tbp]
\epsfxsize=7cm \centerline{\epsffile{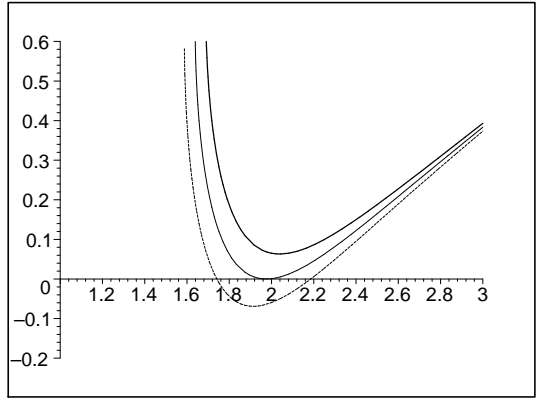}}
\caption{$f(r)$ versus $r$ for $k=0$, $p=1$, $\protect\alpha=0.1$, $%
\Lambda=-1$, $r_{0}=0.1$, $q=2.5$ and $m=43<m_{\mathrm{ext}}$ (bold line), $%
m=45.1381=m_{\mathrm{ext}}$ (continuous line) and
$m=47>m_{\mathrm{ext}}$ (dotted line).} \label{Fig f(r)}
\end{figure}
One can find the extremal value of mass, $m_{\mathrm{ext}}$, in
terms of parameters of metric function by finding $r_{+}$ from Eq.
(\ref{f'}) and
inserting it into equation $f(r_{+})=0$. Then, the metric of Eqs. (\ref%
{metric}), (\ref{F(r)}) and (\ref{g(r)}) presents a black hole
solution with inner and outer event horizons provided
$m>m_{\mathrm{ext}}$, an extreme black hole for
$m=m_{\mathrm{ext}}$ [temperature is zero since it is proportional
to $f^{\prime }(r_{+})$] and a naked singularity otherwise. It is
a matter of calculation to show that $m_{\mathrm{ext}}$ for $k=0$
becomes
\begin{equation}
m_{\mathrm{ext}}=-\frac{4\Lambda \left( p+1\right) }{(2p+1)(4p+3)}\left( -%
\frac{2^{p}\left( 2p+1\right) q_{\mathrm{ext}}^{2p+2}}{\Lambda
}\right) ^{\left( 4p+3\right) /\left( 4p+4\right) }.  \label{mext}
\end{equation}
The Hawking temperature of the black holes can be easily obtained
by requiring the absence of conical singularity at the horizon in
the Euclidean sector of the black hole solutions. One obtains
\begin{eqnarray}
{T}_{+} &=&\frac{f^{\prime }(r_{+})}{4\pi }\sqrt{1+\frac{r_{0}^{2}}{r_{+}^{2}%
}}=  \nonumber \\
&&\frac{(4p+1)k\left[ 4pk(4p-1)\alpha
+(r_{+}^{2}+r_{0}^{2})\right]
(r_{+}^{2}+r_{0}^{2})^{2p}-\frac{\Lambda (r_{+}^{2}+r_{0}^{2})^{2p+2}}{2p+1}%
-2^{p}q^{2p+2}}{4\pi (r_{+}^{2}+r_{0}^{2})^{(4p+1)/2}\left[
r_{+}^{2}+r_{0}^{2}+8pk(4p+1)\alpha \right] }.  \label{Temp}
\end{eqnarray}%
It is worth to note that $T_{+}$ is zero for $m=m_{\mathrm{ext}}$.

\subsection{Thermodynamics of Asymptotically Flat Black Holes for $k=1$\label%
{flat}} In this section, we consider the thermodynamics of
spherically symmetric black holes which are asymptotically flat.
First of all we focus on entropy. Usually entropy of black holes
satisfies the so-called area law of entropy which states that the
black hole entropy equals to one-quarter of horizon area. It
applies to all kind of black holes and black strings of Einstein
gravity \cite{Beck}. However, in higher derivative gravity the
area law of entropy is not satisfied in general \cite{fails}. It
is known that the entropy in of asymptotically flat solutions of
Lovelock gravity is \cite{Myers}
\begin{equation}
S=\frac{1}{4}\sum_{k=1}^{[2p+1/2]}k\alpha _{k}\int d^{4p+2}x\sqrt{\tilde{g}}%
\tilde{\mathcal{L}}_{k-1},  \label{Enta}
\end{equation}%
where the integration is done on the $(4p+2)$-dimensional
spacelike hypersurface of Killing horizon, $\tilde{g}_{\mu \nu }$
is the induced
metric on it, $\tilde{g}$ is the determinant of $\tilde{g}_{\mu \nu }$ and $%
\tilde{\mathcal{L}}_{k}$ is the $k$th order Lovelock Lagrangian of $\tilde{g}%
_{\mu \nu }$. Denoting the volume of the boundary at constant $t$
and $r$ by $V_{4p+2}$, the entropy in Gauss-Bonnet gravity per
unit volume $V_{4p+2}$ is
\begin{equation}
S=\frac{1}{4}\int d^{4p+2}x\sqrt{\tilde{g}}\left( 1+2\alpha
\tilde{R}\right) ,  \label{Entb}
\end{equation}%
where $\tilde{R}$ is the Ricci scalar for the induced metric
$\tilde{g}_{ab}$ on the $(4p+2)$-dimensional horizon. It is a
matter of calculation to show that the entropy of black holes is
\begin{equation}
S=\frac{(r_{+}^{2}+r_{0}^{2})^{2p+1}}{4}\left( 1+\frac{4(4p+1)(2p+1)\alpha }{%
r_{+}^{2}+r_{0}^{2}}\right) .  \label{Ent1}
\end{equation}%
The charge per unit volume $V_{4p+2}$ of the black hole can be
found by calculating the flux of the electric field at infinity,
yielding
\begin{equation}
Q=\frac{2^{p}(p+1)q^{2p+1}}{4\pi }.  \label{Ch}
\end{equation}%
The electric potential $\Phi $, measured at infinity with respect
to the horizon, is defined by
\begin{equation}
\Phi =A_{\mu }\chi ^{\mu }\left\vert _{r\rightarrow \infty
}-A_{\mu }\chi ^{\mu }\right\vert _{r=r_{+}},  \label{Pot1}
\end{equation}%
where $\chi =\partial /\partial t$ is the null generator of the
horizon. One finds
\begin{equation}
\Phi =\frac{q}{\sqrt{r_{+}^{2}+r_{0}^{2}}}.  \label{Pot2}
\end{equation}%
The ADM (Arnowitt--Deser--Misner) mass of black hole can be
obtained by using the behavior of the metric at large $r$.
According to the definition of mass due to Abbott and Deser
\cite{AbbDes}, the mass per unit volume $V_{4p+2}$ of the
solutions ($k=1$) is
\begin{equation}
M=\frac{1}{8\pi }\left( 2p+1\right) m.  \label{Mass1}
\end{equation}
We now investigate the first law of thermodynamics. Using the
expression for the entropy, the charge, and the mass given in Eqs.
(\ref{Ent1}), (\ref{Ch}) and (\ref{Mass1}), and the fact that
$f(r_{+})=0$, one obtains
\begin{eqnarray}
M(S,Q) &=&\frac{2p+1}{8\pi \sqrt{r_{+}^{2}+r_{0}^{2}}}\left[
\left( r_{+}^{2}+r_{0}^{2}\right) ^{2p+1}+4p\alpha (4p+1)\left(
r_{+}^{2}+r_{0}^{2}\right) ^{2p}-\frac{\Lambda \left(
r_{+}^{2}+r_{0}^{2}\right) ^{2p+2}}{(2p+1)(4p+3)}+\right.  \nonumber \\
&&\left. 2^{p}\left( \frac{4\pi Q}{2^{p}(p+1)}\right)
^{2(p+1)/\left( 2p+1\right) }\right] ,  \label{Sma1}
\end{eqnarray}%
where $r_{+}$\ is the real root of Eq. (\ref{Ent1}) which is a function of $%
S $. One may then regard the parameters $S$ and $Q$ as a complete
set of extensive parameters for the mass $M(S,Q)$ and define the
intensive parameters conjugate to them. These quantities are the
temperature and the electric potential
\begin{equation}
T=\left( \frac{\partial M}{\partial S}\right) _{Q},\ \ \ \ \Phi
=\left( \frac{\partial M}{\partial Q}\right) _{S}.  \label{Dsma1}
\end{equation}%
Computing $\partial M/\partial r_{+}$ and $\partial S/\partial
r_{+}$ and using chain rule, it is easy to show that the intensive
quantities
calculated by Eq. (\ref{Dsma1}) coincide with Eqs. (\ref{Temp}) and (\ref%
{Pot2}) respectively. Thus, the thermodynamic quantities calculated in Eqs. (%
\ref{Temp}) and (\ref{Pot2}) satisfy the first law of
thermodynamics,
\begin{equation}
dM=TdS+\Phi dQ.  \label{1stlaw}
\end{equation}

\subsection{Thermodynamics of Asymptotically AdS Rotating Black Branes with
Flat Horizon\label{rotating}} Now, we want to endow our spacetime
solution (\ref{met}) for $k=0$\ with a global rotation. In order
to add angular momentum to the spacetime, we perform the following
rotation boost in the $t-\phi _{i}$ planes
\begin{equation}
t\mapsto \Xi t-a_{i}\phi _{i},\hspace{0.5cm}\phi _{i}\mapsto \Xi \phi _{i}-%
\frac{a_{i}}{l^{2}}t,  \label{Tr}
\end{equation}%
for $i=1...[(4p+3)/2]$, where $[x]$ is the integer part of $x$.
The maximum
number of rotation parameters is due to the fact that the rotation group in $%
4p+4$ dimensions is $SO(4p+3)$ and therefore the number of
independent rotation parameters is $[(4p+3)/2]$. Thus the metric
of asymptotically AdS rotating solution with $n\leq \lbrack
(4p+3)/2]$ rotation parameters for flat horizon can be written as
\begin{eqnarray}
ds^{2} &=&-f(r)\left( \Xi dt-{{\sum_{i=1}^{n}}}a_{i}d\phi _{i}\right) ^{2}+%
\frac{r^{2}+r_{0}^{2}}{l^{4}}{{\sum_{i=1}^{n}}}\left( a_{i}dt-\Xi
l^{2}d\phi
_{i}\right) ^{2}  \nonumber \\
&&\ \text{ }+\frac{r^{2}dr^{2}}{(r^{2}+r_{0}^{2})f(r)}-\frac{r^{2}+r_{0}^{2}%
}{l^{2}}{\sum_{i<j}^{n}}(a_{i}d\phi _{j}-a_{j}d\phi
_{i})^{2}+\left( r^{2}+r_{0}^{2}\right)
{{\sum_{i=n+1}^{4p+2}}}d\phi _{i},  \label{met1}
\end{eqnarray}%
where $\Xi =\sqrt{1+\sum_{i}^{n}a_{i}^{2}/l^{2}}$. Because of the
periodic nature of $\phi$, the transformation \ref{Tr} is not a
proper coordinate transformation on the entire manifold.
Therefore, the static and rotating metrics can be locally mapped
into each other but not globally, and so they are distinct
\cite{stachel}. We show that the intensive and extensive
quantities depend on the rotation parameters. Using Eq.
(\ref{Maxeq}), one can show that the vector potential can be
written as
\begin{equation}
A_{\mu }=-\frac{q}{\sqrt{r^{2}+r_{0}^{2}}}\left( \Xi \delta _{\mu
}^{0}-\delta _{\mu }^{i}a_{i}\right) \text{(no sum on }i\text{).}
\label{Amu}
\end{equation}%
One can obtain the temperature and angular momentum of the event
horizon by analytic continuation of the metric. One obtains
\begin{equation}
{T}_{+}{=}\frac{f^{\prime }(r_{+})}{4\pi \Xi }\sqrt{1+\frac{r_{0}^{2}}{%
r_{+}^{2}}}=-\frac{\Lambda (r_{+}^{2}+r_{0}^{2})^{2p+2}+2^{p}(2p+1)q^{2p+2}}{%
4\pi \Xi (2p+1)(r_{+}^{2}+r_{0}^{2})^{(4p+3)/2}},  \label{Tempk0}
\end{equation}%
\begin{equation}
\Omega _{i}=\frac{a_{i}}{\Xi l^{2}}.  \label{Om}
\end{equation}%
Next, we calculate the electric charge and potential of the
solutions. The electric charge per unit volume $V_{4p+2}$ can be
found by calculating the flux of the electric field at infinity,
yielding
\begin{equation}
Q=\frac{2^{p}(p+1)\Xi q^{2p+1}}{4\pi }.  \label{Charge}
\end{equation}%
Using Eq. (\ref{Pot1}) and the fact that $\chi =\partial _{t}+{\sum_{i}^{n}}%
\Omega _{i}\partial _{\phi _{i}}$ is the null generator of the
horizon, the electric potential $\Phi $ is obtained as
\begin{equation}
\Phi =\frac{q}{\Xi \sqrt{r_{+}^{2}+r_{0}^{2}}}.  \label{Pot}
\end{equation}

\subsubsection{Finite Action and Conserved quantities of the solutions}
Here, we calculate the action and conserved quantities of the
black brane solutions. For asymptotically AdS of our solutions with flat boundary, $%
\widehat{R}_{abcd}(\gamma )=0$, the finite action, $I_{G}+I_{ct}$,
reduce to
\cite{DehBord,DM1}%
\begin{equation}
I=I_{G}+\frac{1}{4\pi }\int_{\partial \mathcal{M}}d^{4p+3}x\sqrt{-\gamma }%
\left( \frac{2p+1}{L}\right) ,  \label{Ifinite}
\end{equation}%
where $L$\ is
\begin{eqnarray}
L &=&\frac{3l\sqrt{\zeta l\left( l-\sqrt{l^{2}-2\zeta }\right) }}{%
l^{2}+2\zeta -l\sqrt{l^{2}-2\zeta }},  \nonumber \\
\zeta &=&8p(4p+1)\alpha .  \label{L}
\end{eqnarray}%
One may note that $L$ reduces to $l$\ as $\alpha $\ goes to zero.
Using Eqs. (\ref{Ig}) and (\ref{Ifinite}), the finite action per
unit volume $V_{4p+2}$ can be calculated as
\begin{equation}
I=\frac{1}{16\pi \sqrt{r_{+}^{2}+r_{0}^{2}}l^{2}T_{+}}\left[ 2(2p+1)l^{2}m%
\sqrt{r_{+}^{2}+r_{0}^{2}}-2^{p}(4p+3)l^{2}q^{2p+2}-{%
(4p+3)(r_{+}^{2}+r_{0}^{2})}^{{2p+2}}\right] .  \label{finiteAct}
\end{equation}%
Using the Brown-York method \cite{BY}, the finite energy-momentum
tensor is
\begin{equation}
T^{ab}=\frac{1}{8\pi }\{(K^{ab}-K\gamma ^{ab})+2\alpha
(3J^{ab}-J\gamma ^{ab})+\frac{2\left( 2p+1\right) }{L}\gamma
^{ab}\ \},  \label{Tab}
\end{equation}%
and by using Eq. (\ref{quasi}), the conserved quantities
associated with the Killing vectors $\partial /\partial t$ and
$\partial /\partial \phi ^{i}$ are
\begin{eqnarray}
M &=&\frac{1}{16\pi }\left[ (4p+3)\Xi ^{2}-1\right] m,  \label{Mass} \\
J_{i} &=&\frac{1}{16\pi }(4p+3)\Xi ma_{i},  \label{Angmom}
\end{eqnarray}%
which are the mass and angular momentum per unit volume $V_{4p+2}$
of the solutions.

Now using Gibbs-Duhem relation
\begin{equation}
S=\frac{1}{T}(M-Q\Phi -{{\sum_{i=1}^{n}}}\Omega _{i}J_{i})-I,
\label{GibsDuh}
\end{equation}%
and Eqs. (\ref{Pot}), (\ref{finiteAct}), (\ref{Mass}) and
(\ref{Angmom}), one obtains
\begin{equation}
S=\frac{\Xi }{4}\left( r_{+}^{2}+r_{0}^{2}\right) ^{2p+1},
\label{Entropy}
\end{equation}%
for the entropy per unit volume $V_{4p+2}$. This shows that the
entropy obeys the area law for recent case where the horizon is
flat.

\subsubsection{First law of thermodynamics \ and Stability of the solutions
\label{Stab}} Calculating all the thermodynamic and conserved
quantities of the black brane solutions, we now check the first
law of thermodynamics for our solutions with flat horizon. We
obtain the mass as a function of the extensive quantities $S$,
$\mathbf{J}$, and $Q$. Using the expression for
charge mass, angular momenta and entropy given in Eqs. (\ref{Charge}), (\ref%
{Mass}), (\ref{Angmom}), (\ref{Entropy}) and the fact that
$f(r_{+})=0$, one can obtain a Smarr-type formula as
\begin{equation}
M(S,\mathbf{J},Q)=\frac{\left[ \left( 4p+3\right) Z-1\right]
J}{\left( 4p+3\right) l\sqrt{Z(Z-1)}},  \label{Smar}
\end{equation}%
where $J=\left\vert \mathbf{J}\right\vert
=\sqrt{\sum_{i}^{n}J_{i}^{2}}$ and
$Z=\Xi ^{2}$ is the positive real root of the following equation%
\begin{equation}
S^{(4p+3)/(4p+2)}+\frac{\pi Ql^{2}}{p+1}\left( \frac{\pi ^{2}Q^{2}}{%
2^{2p}\left( p+1\right) ^{2}S}\right)
^{1/(4p+2)}-\frac{2^{2p/(2p+1)}\pi
lJZ^{1/(8p+4)}}{(4p+3)\sqrt{Z-1}}=0.
\end{equation}%
One may then regard the parameters $S$, $J_{i}$'s, and $Q$ as a
complete set of extensive parameters for the mass
$M(S,\mathbf{J},Q)$ and define the intensive parameters conjugate
to them. These quantities are the temperature, the angular
velocities, and the electric potential
\begin{equation}
T=\left( \frac{\partial M}{\partial S}\right) _{J,Q},\ \ \Omega
_{i}=\left(
\frac{\partial M}{\partial J_{i}}\right) _{S,Q},\ \ \Phi =\left( \frac{%
\partial M}{\partial Q}\right) _{S,J}.  \label{Dsmar}
\end{equation}%
Straightforward calculations show that the intensive quantities
calculated
by Eq. (\ref{Dsmar}) coincide with Eqs. (\ref{Tempk0}), (\ref{Om}) and (\ref%
{Pot}). Thus, these quantities satisfy the first law of
thermodynamics
\[
dM=TdS+{{{\sum_{i=1}^{n}}}}\Omega _{i}dJ_{i}+\Phi dQ.
\]

\begin{figure}[tbp]
\epsfxsize=7cm \centerline{\epsffile{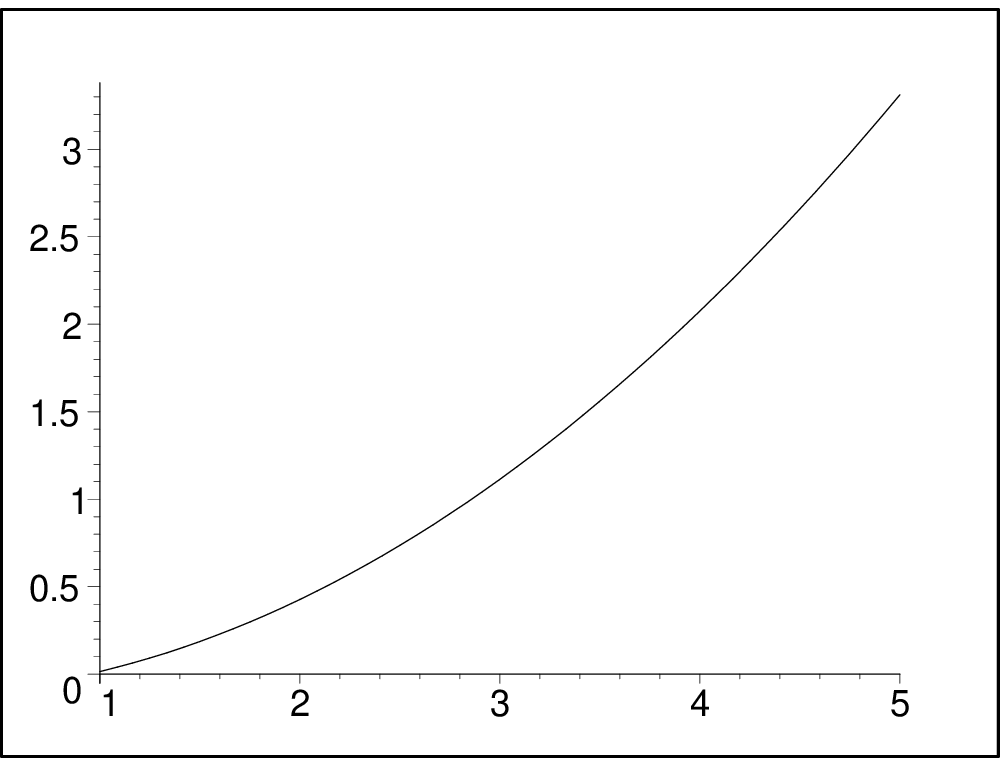}}
\caption{$10^{-13}\Upsilon $ versus $\Xi $ for $r_{0}=3$,
$r_{+}=4$, $l=0.1$, $q=0.1$ and $p=1$.} \label{Fig op1}
\end{figure}
\begin{figure}[tbp]
\epsfxsize=7cm \centerline{\epsffile{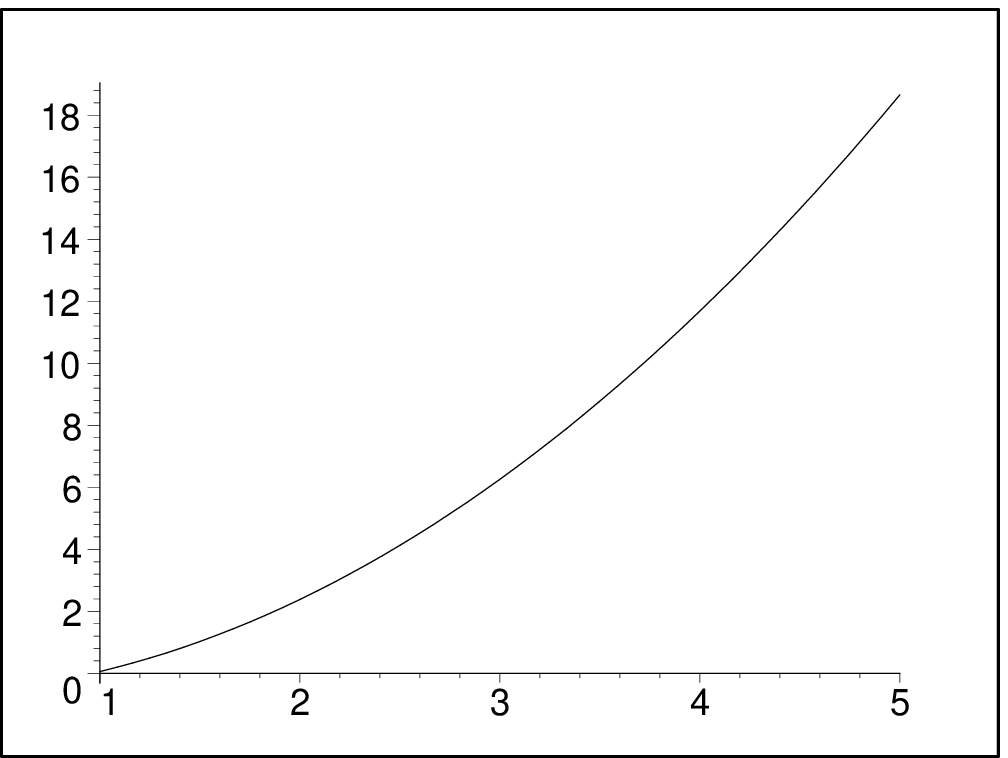}}
\caption{$10^{-18}\Upsilon $ versus $\Xi $ for $r_{0}=3$,
$r_{+}=4$, $l=0.1$, $q=0.1$ and $p=2$.} \label{Fig op2}
\end{figure}
\begin{figure}[tbp]
\epsfxsize=7cm \centerline{\epsffile{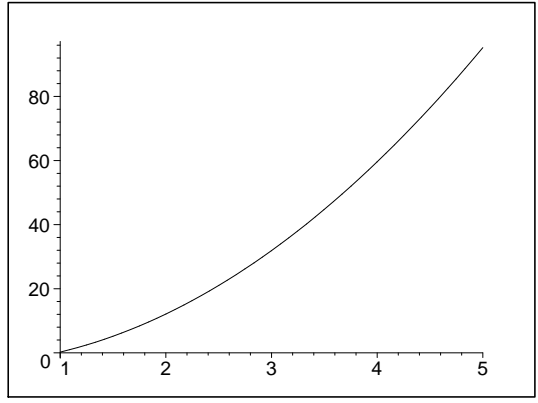}}
\caption{$10^{-23}\Upsilon $ versus $\Xi $ for $r_{0}=3$,
$r_{+}=4$, $l=0.1$, $q=0.1$ and $p=3$.} \label{Fig op3}
\end{figure}
Finally, we investigate the local stability of charged rotating
black brane solutions of Gauss-Bonnet gravity in the presence of
conformally invariant Maxwell source in the canonical and grand
canonical ensembles. In the
canonical ensemble, the positivity of the heat capacity $C_{\mathbf{J}%
,Q}=T_{+}/(\partial ^{2}M/\partial S^{2})_{\mathbf{J},Q}$ and
therefore the positivity of $(\partial ^{2}M/\partial
S^{2})_{\mathbf{J},Q}$ is sufficient to ensure the local
stability. It is easy show that
\begin{equation}
\frac{\partial ^{2}M}{\partial S^{2}}=\frac{\left( 4p+3\right)
\left[
(r_{+}^{2}+r_{0}^{2})^{2p+2}+2^{p}q^{2\left( p+1\right) }l^{2}\right] ^{-1}}{%
\pi \Xi ^{2}l^{2}\left[ (p+1)\Xi ^{2}+1\right] (r_{+}^{2}+r_{0}^{2})^{4p+5/2}%
}\Upsilon ,  \label{dMSS}
\end{equation}%
where
\begin{eqnarray}
\Upsilon &=&2^{2p}l^{4}\Xi ^{2}q^{4(p+1)}+\frac{2^{p+1}}{(2p+1)}%
q^{2(p+1)}l^{2}(r_{+}^{2}+r_{0}^{2})^{2p+2}[2(p+1)-\Xi ^{2}]+ \\
&&[4(p+1)\left( \Xi ^{2}-1\right) +\Xi
^{2}](r_{+}^{2}+r_{0}^{2})^{4p+4} \nonumber
\end{eqnarray}
The heat capacity is positive for $m\geq m_{\mathrm{ext}}$, where
the temperature is positive. This fact can be seen easily for $\Xi
=1$, where
the $\Upsilon $ is positive and then $(\partial ^{2}M/\partial S^{2})_{%
\mathbf{J},Q}$ is positive too. Also, one may see from Figs. (\ref{Fig op1}, %
\ref{Fig op2} and \ref{Fig op3}) that the $\Upsilon $ increases as
$\Xi $ increases, and therefore it is always positive. Thus the
condition for thermal equilibrium in the canonical ensemble is
satisfied.

In grand canonical ensemble, the positivity of the determinant of
Hessian matrix of $M(S,Q,\mathbf{J})$ with respect to its extensive variables $X_{i}$%
, $\mathbf{H}_{X_{i}X_{j}}^{M}=\left( \partial ^{2}M/\partial
X_{i}\partial X_{j}\right) $, is sufficient to ensure the local
stability. It is a matter
of calculation to show that the determinant of $\mathbf{H}_{S,Q,\mathbf{J}%
}^{M}$ is%
\begin{equation}
\left\vert \mathbf{H}_{SJQ}^{M}\right\vert =\frac{64\pi \left[
\left( 4p+3\right) (r_{+}^{2}+r_{0}^{2})^{2p+2}+2^{p}\left(
2p+1\right) l^{2}q^{2\left( p+1\right) }\right] \left[
(r_{+}^{2}+r_{0}^{2})^{2p+2}+2^{p}l^{2}q^{2\left( p+1\right) }\right] ^{-1}}{%
\left[ (4p+1)\Xi ^{2}+1\right] \left[ 2^{p}\left( 4p+3\right)
\left( 2p+1\right) (p+1)l^{2}\Xi
^{6}q^{2p}(r_{+}^{2}+r_{0}^{2})^{4p+5/2}\right] }. \label{Hes}
\end{equation}%
Equation (\ref{Hes}) shows that the determinant of Hessian matrix
is positive, and therefore the solutions are stable in grand
canonical ensemble too. This phase behavior is commensurate with
the fact that there is no Hawking-Page transition for a black
object whose horizon is diffeomorphic to $\mathbb{R}^{p}$ and
therefore the system is always in the high temperature phase
\cite{Wit2}.

\section{ Closing Remarks}
In this paper, we presented a class of rotating solutions in
Gauss--Bonnet gravity in the presence of conformally invariant
Maxwell field and investigated the effects of the nonlinearity of
the electromagnetic fields on the properties of the solutions.
These solutions may be interpreted as black brane solutions with
inner and outer event horizons provided that the mass parameter
$m$ is greater than an extremal value $m_{ext}$, an extreme black
brane if $m=m_{ext}$ and a naked singularity otherwise. We
considered thermodynamics of asymptotically flat solutions and
found that the first law of thermodynamics is satisfied by the
conserved and thermodynamic quantities of the black hole. We also
considered the rotating solution with flat horizon and computed
the action and conserved quantities of it through the use of
counterterm method. We found that the entropy obeys the area law
for black branes with flat horizon. We obtained a Smarr-type
formula for the mass of the black brane as a function of the
entropy, the charge, and the angular momenta, and found that the
conserved and thermodynamics quantities satisfy the first law of
thermodynamics. We also studied the phase behavior of the
($n+1$)-dimensional rotating black branes and showed that there is
no Hawking-Page phase transition in spite of the angular momenta
of the branes and the presence of the conformally invariant
Maxwell field. Indeed, we calculated the heat capacity and the
determinant of the Hessian matrix of the mass with respect to $S$,
$\textbf{J}$, and $Q$ of the black branes and found that they are
positive for all the phase space, which means that the brane is
locally stable for all the allowed values of the metric
parameters.
\begin{acknowledgements}
This work has been supported financially by Research Institute for
Astronomy and Astrophysics of Maragha.
\end{acknowledgements}


\begin{thebibliography}{99}
\bibitem{Birrel} N. D. Birrell and P. C. W. Davies, Quantum Fields in Curved
Space, (Cambridge University Press, Cambridge, England, 1982).

\bibitem{Green} M. B. Greens, J. H. Schwarz and E. Witten, Superstring
Theory, (Cambridge University Press, Cambridge, England, 1987); D.
Lust and S. Theusen, Lectures on String Theory, (Springer, Berlin,
1989); J. Polchinski, String Theory, (Cambridge University Press,
Cambridge, England, (1998).

\bibitem{Lovelock} D. Lovelock, J. Math. Phys. \textbf{12}, 498 (1971); N.
Deruelle and L. Farina-Busto, Phys. Rev. D \textbf{41}, 3696
(1990); G. A. MenaMarugan, Phys. Rev. D \textbf{46}, 4320 (1992);
4340 (1992).

\bibitem{Boulware} D. G. Boulware and S. Deser, Phys. Rev. Lett. \textbf{55}%
, 2656 (1985); J. T. Wheeler, Nucl. Phys. B \textbf{268}, 737
(1986).

\bibitem{Astefanesei} D. Astefanesei, N. Banerjee and S. Dutta JHEP 0811, 070 (2008).

\bibitem{Wil1} D. L. Wiltshire, Phys. Lett. B \textbf{169}, 36 (1986); D. L. Wiltshire, Phys.
Rev. D \textbf{38}, 2445 (1988).

\bibitem{DH} M. H. Dehghani and S. H. Hendi, Int. J. Mod. Phys. D \textbf{16}%
, 1829 (2007).

\bibitem{Deh1} M. H. Dehghani, Phys. Rev. D \textbf{67}, 064017 (2003).

\bibitem{Deh2} M. H. Dehghani, Phys. Rev. D \textbf{70}, 064019 (2004).

\bibitem{DehMagGB} M. H. Dehghani, Phys. Rev. D 69 064024 (2004).

\bibitem{Deh3} M. H. Dehghani and M. Shamirzaie, Phys. Rev. D \textbf{72},
124015 (2005).

\bibitem{MS} R. C. Myers and J. Z. Simon, Phys. Rev. D \textbf{38}, 2434
(1988); R. C. Myers, Nucl. Phys. B \textbf{289}, 701 (1987).

\bibitem{Cai} R. G. Cai, Phys. Rev. D \textbf{65}, 084014 (2002).

\bibitem{NUT} M. H. Dehghani and R. B. Mann, Phys. Rev. D \textbf{72},
124006 (2005); M. H. Dehghani and S. H. Hendi, Phys. Rev. D 73,
084021 (2006).

\bibitem{Od1} M. Cvetic, S. Nojiri and S. D. Odintsov, Nucl. Phys. B \textbf{%
628}, 295 (2002); S. Nojiri and S. D. Odintsov, Phys. Lett.
\textbf{521B}, 87 (2001).

\bibitem{DesSch} S. Deser and A. Schwimmer, Int. J. Mod. Phys. B 8, 3741
(1994).

\bibitem{EspSto} G. Esposito and C. Stornaiolo, Class. Quant. Grav. 17, 1989
(2000); G. Esposito and C. Stornaiolo, Nucl. Phys. Proc. Suppl.
88, 365 (2000)

\bibitem{CodOsb} C. Codirla and H. Osborn, Annals Phys. 260, 91 (1997).

\bibitem{EasSin} M. Eastwood and M. Singer, Phys. Lett. A 107, 73 (1985).

\bibitem{Pop} N. J. Poplawski, Int. J. Mod. Phys. A 23, 567 (2008).

\bibitem{Esp} G. Esposito, Phys. Rev. D56, 2442 (1997).

\bibitem{HasMar} M. Hassaine and C. Martinez, Phys. Rev. D. 75, 027502
(2007); M. Hassaine and C. Martinez, Class. Quant. Grav. 25,
195023 (2008); H. Maeda, M. Hassaine and C. Martinez, Phys. Rev. D
79, 044012 (2009).

\bibitem{MyeDavis} R. C. Myers, Phys. Rev. D \textbf{36}, 392 (1987); S. C.
Davis, Phys. Rev. D \textbf{67}, 024030 (2003).

\bibitem{Mal} J. Maldacena, Adv. Theor. Math. Phys., \textbf{2}, 231 (1998);
E. Witten, Adv. Theor. Math. Phys., \textbf{2}, 253 (1998); O.
Aharony, S. S. Gubser, J. Maldacena, H. Ooguri and Y. Oz, Phys.
Rept., \textbf{323}, 183 (2000).

\bibitem{Deh33} M. H. Dehghani, Phys. Rev. D \textbf{66}, 044006 (2002);
Phys. Rev. D \textbf{65}, 124002 (2002); M. H. Dehghani and A.
Khodam-Mohammadi, Phys. Rev. D 67, 084006 (2003).

\bibitem{DM1} M. H. Dehghani and R. B. Mann, Phys. Rev. D \textbf{64},
044003 (2001); M. H. Dehghani, Phys. Rev. D \textbf{65}, 104030
(2002); M. H. Dehghani and H. KhajehAzad, Can. J. Phys.
\textbf{81}, 1363 (2003).

\bibitem{Kraus} P. Kraus, F. Larsen and R. Siebelink, Nucl. Phys. B \textbf{%
563}, 259 (1999).

\bibitem{BY} J. D. Brown and J. W. York, Phys. Rev. D\textbf{\ 47}, 1407
(1993).

\bibitem{DehBord} M. H. Dehghani, G. H. Bordbar and M. Shamirzaie, Phys.
Rev. D \textbf{74}, 064023 (2006).

\bibitem{Beck} J. D. Bekenstein, Phys. Rev. D \textbf{7}, 2333 (1973); S. W.
Hawking and C. J. Hunter, Phys. Rev. D \textbf{59} 044025 (1999);
S. W. Hawking, C. J. Hunter and D. N. Page, Phys. Rev. D
\textbf{59}, 044033 (1999); R. B. Mann, Phys. Rev. D \textbf{60},
104047 (1999); Phys. Rev. D \textbf{61}, 084013 (2000); C. J.
Hunter, Phys. Rev. D \textbf{59}, 024009 (1999).

\bibitem{fails} M. Lu and M. B. Wise, Phys. Rev. D \textbf{47}, 3095 (1993);
M. Visser, Phys. Rev. D \textbf{48}, 583 (1993).

\bibitem{Myers} T. Jacobson and R. C. Myers, Phys. Rev. Lett. \textbf{70},
3684 (1993); R. M. Wald, Phys. Rev. D \textbf{48}, R3427, (1993);
M. Visser, Phys. Rev. D \textbf{48}, 5697 (1993); T. Jacobson, G.
Kang and R. C. Myers, Phys. Rev. D \textbf{49}, 6587,(1994); V.
Iyer and R. M. Wald, Phys. Rev. D \textbf{50}, 846 (1994).

\bibitem{AbbDes} L. F. Abbott and S. Deser, Nucl. Phys. B \textbf{195}, 76
(1982).

\bibitem{stachel} J. Stachel, Phys. Rev. D 26, 1281 (1982).

\bibitem{Wit2} E. Witten, Adv. Theor. Math. Phys. \textbf{2}, 505 (1998).

\end{thebibliography}
\end{document}